\documentclass[12pt]{article}
\usepackage{aaspp4}
\usepackage{epsfig}

\begin{document}

\title{The Spectrum of Diffuse Cosmic Hard X-Rays Measured with HEAO--1}
\author{D. E. Gruber, J. L. Matteson and L. E. Peterson}
\affil{Center for Astrophysics \& Space Sciences, University of California,
San Diego 92093}
\authoremail{dgruber@ucsd.edu}
\author{G. V. Jung}
\affil{Naval Research Laboratory, Washington, DC  20375}

\begin{abstract}
The spectrum of the diffuse isotropic component of 
cosmic X-rays over the 13--180 keV range was
determined by the UCSD/MIT Hard X-Ray and Gamma-Ray instrument (HEAO A4)
on the High Energy Astronomical Observatory--1 (HEAO--1).  
The instrument consists of a complex of actively 
shielded and collimated 
scintillation counters, including the Low Energy Detector 
set from which the data reported here were obtained.  These data 
join smoothly with the spectrum at lower energies reported by the 
GSFC HEAO A2 instrument and with that measured to 400 keV by the 
HEAO A4 Medium Energy Detectors.  The HEAO data set also joins 
the recent results from COMPTEL on the Compton Gamma-Ray Observatory in the 
1--10 MeV range, which failed 
to confirm the existence of an ``MeV bump'' in this range.
Although the spectrum over the entire range 
3 keV $\leq$ E $\leq$ 100 GeV can be fit by a simple empirical 
analytic expression,
the origin is likely due to a number of distinct source components.
The prevailing idea for the origin is that the hard 
X-ray spectrum is due to X-rays from various AGN components, 
particularly Seyfert galaxies extending to 
cosmological 
distances, and that the low energy gamma-rays may be due to emission
from type 1a Supernovae, also integrated to cosmological distances.
The higher energy gamma-ray spectrum defined by EGRET, also on the 
CGRO, may be due to unresolved gamma-ray emitting blazars.
Models of production by these source components, extrapolated to the
present epoch, must reproduce the observationally derived spectrum.
\end{abstract}


\section{Introduction}

The spectrum of the diffuse sky background of cosmic X- and gamma-rays has
been a matter of considerable interest and some controversy since the discovery
by rocket-borne X-ray counters (Giacconi 1962) and by a gamma-ray counter
on the Ranger III lunar probe (Metzger et al. 1968).
The known spectrum was extended beyond 100 MeV by an instrument on OSO-III
(Kraushaar et al. 1972).
Although there were many 
subsequent measurements by a variety of rocket, balloon and space-borne 
instruments during the 1960's and early 1970's (Horstman et al. 1975), 
the most definitive spectra below about 500 keV were obtained from the
HEAO--1, launched in 1977 (Marshall et al. 1980; Kinzer et al.
1997).  

At higher energies (i.e. $>$ 800 keV) the spectrum 
has recently been clarified with data obtained from the Compton Gamma-Ray
Observatory (CGRO).  The COMPTEL instrument on the CGRO 
has failed to confirm the ``MeV 
Bump'' (Trombka et al. 1977)
in the diffuse gamma-ray spectrum in the range 0.8$\leq$ E $\leq$ 30 MeV
(Kappadath et al. 1995, 1996), while the EGRET instrument 
(Kniffen et al 1996, Sreekumar et al. 1998) has generally confirmed the 
results presented earlier 
by Fichtel in the 100 MeV range by a spark chamber on the Small Astronomy
Satellite--2 (SAS--2) (Fichtel et al. 1978), and has also extended the spectrum
to about 100 GeV.

The near isotropy of the diffuse X-ray background and its large energy density 
point to an
extragalactic and even cosmological origin.  Early attempts to produce the
spectrum above about 3 keV in terms of uniform emissions at truly 
cosmological distances seem to have been ruled out (Barcons, Fabian \& Rees 
1991); therefore discrete source populations which extends to high 
redshifts must be considered  (Barber \& Warwick 1994).  Fabian and Barcons 
(1992) and Hasinger (1996) provide reviews of the observational 
and theoretical status
of the subject as of these dates.  
The most recent concept, summarized by Zdziarski (1996) is 
that the background in the range of $\sim$3--300 keV is due to various 
AGN components, particularly Seyfert II's, (Madau et al. 1994), 
and that the low energy gamma-ray background ($\sim$ 300 keV 
$<$E$<$ 10 MeV) is due
to supernova 1a (The, Leising \& Clayton 1993).  The diffuse component 
at energies 
$>$ 30 MeV measured by EGRET is attributed to 
unresolved blazars (Stecker \& Salmon, 1996).  The components in the range
0.4--10 keV, as determined with ASCA (Gendreau et al. 1995), may also be 
accounted for in terms of AGN's; however, there exists an excess below 1 keV 
which, if not accounted for by effects in the local ISM, requires 
additional source components (Chen, Fabian \& Gendreau 1997).
The ROSAT deep X-ray survey in the Lockman hole
has discovered enough sources to account for at least
70--80\% of the diffuse flux in the 0.5--2 keV range (Hasinger 1998).
Taking into account evolution, such as that which characterizes quasars, 
this source density actually overproduces the X-ray background 
in this range (Hasinger, 
pvt comm).

This paper describes the final spectral results obtained by one of the UCSD/MIT
Hard X-ray detectors on the HEAO--1 over the 13--180 keV range.  
This data is compared with related data on the diffuse component,
and the total spectrum to 100 GeV is fit by a simple analytic function.
Preliminary results of this work have been reported earlier 
(Rothschild et al. 1983; Gruber 1992). 

\section{Instrument and Operation}

The UCSD/MIT Hard X-ray and Gamma-Ray Instrument launched on the HEAO--1 
spacecraft has been described previously (Matteson 1978,
Jung 1989, Kinzer et al. 1997).  The instrument consists of an array 
of seven NaI/CsI ``phoswich'' detectors collimated with 
a thick CsI active 
anticoincidence shield. Three different detector configurations were 
optimized to cover different sub-ranges over the 13 keV -- 10 MeV 
total range of the 
instrument.  Relevant properties of the various detectors are indicated in 
Table 1.  The data reported here were taken from one of the two lower energy 
detectors (LED's) which operated over a nominal 13--180 keV energy range.
The LED's had a passive lead-tin-copper multiple-slat subcollimator within 
the circular active CsI aperture to 
give a 1.4$^\circ$~x~20$^\circ$  FWHM beam response.

The HEAO--1 was launched 1977 August 12 into a 22.7$^\circ$, 400 km circular
orbit.  The spacecraft rotated about the Earth-Sun line
with a nominal 33 minute period. 
The detector fields were centered perpendicular to this line, and thus 
scanned across the sky and the Earth below every rotation, and made a complete 
sky scan every 6 months.  The mission produced usable data until 1979 January 
13, and the spacecraft re-entered the atmosphere on 1979 March 5.

``Good'' events, which met the criteria of no detectable energy losses in the 
CsI(Na) part of the phoswich and no anticoincidence shield event above
$\simeq$ 50 keV were coded in a 128 channel pulse height analyzer and 
transmitted in an event-by-event manner, with detector identifications, time 
tag, and dead time information.  Auxiliary information on counting rates of the 
various functions, and housekeeping information were also transmitted.  
Commandable data modes allowed various diagnostics to be sent with each event.

Separating diffuse fluxes from various background effects requires a more
sophisticated instrumental and data analysis approach then that for localized 
sources.  A movable 20.0 cm dia. x 5.0 cm thick 
CsI (Na) blocking crystal was arranged
to cover the various apertures so that intrinsic detector backgrounds could
be separated from fluxes entering the apertures.  The blocking crystal or 
``shutter'' could be operated in 
a ``passive ''or ``active'' mode by using or ignoring the telemetered
anticoincidence information during data analysis.  This allowed determination 
of second-order effects due to radiations from the blocking crystal.  The 
analysis of the various background effects has been described in detail in 
the previous paper reporting on the results of the diffuse cosmic flux in the 
$\sim$ 80--400 keV range (Kinzer et al. 1997) obtained with two of the
Medium Energy Detectors.

\section{Observations and Data Selection}

One of the two Low Energy 
Detectors, LED \#6 (Kinzer et al. 1997), 
was covered on 14 occasions by the blocking crystal for
intervals of six to twelve hours each
between 1978 November and 1979 January.  Given losses from live time
correction and incomplete data recovery, this resulted in 205 ks of 
observation with the aperture closed, therefore counting only the cosmic-ray
induced internal background.  The observing intervals were selected
to avoid passage through the South Atlantic Anomaly (SAA) region
of geomagnetically-trapped particles, which also induce sizeable internal
background, most notably from the production of I$^{128}$, which decays 
with a 28 minute half life (Gruber, Jung \& Matteson 1989;
Briggs 1992).  To further avoid this induced background component, 
observations were initiated at least three hours after the last of the 
daily sequence of passages through the SAA. This selection of orbits
resulted, however, in a wide range of geomagnetic cutoffs, 
and therefore of 
fluxes due to cosmic rays and their atmospheric and spacecraft secondaries.
Nevertheless, uniform sampling of the geomagnetic coordinate space B (magnetic 
field) and L (earth radii) (McIlwain, 1961)
was assured by 
the large elapsed time, about 46 orbits.
The closed aperture background was averaged over all zenith angles, since 
effects due to varying aspects of the internal background are expected to be 
very small.

Sky-looking data during this period totaling 224 ks was 
also selected for the
same geomagnetic conditions (B,L) during the SAA-quiet part of the observing
day.  The average of the geomagnetic parameter L agreed within 0.2 percent
of that obtained during the orbits with aperture blocked.  Data were
obtained during scanning observations on a set of sky great circles whose
center moved with the sun during this interval from 16h to 19h R.A.  and
with an average declination of -22 degrees.  A small fraction of data containing
catalogued sources (Levine et al. 1984) was excluded.

\section{Control of Systematics}

\subsection{Variation of Detector Internal Background}

While the counting rate from the diffuse background dominates the
internal background at energies from threshold to $\sim$20 keV, the
sky contribution to the total drops rapidly with energy to 
a few percent at 100 keV.  Since the internal background varies with
geomagnetic L to a power between 1 and 2 (Gruber 1974), 
the average of L to $<$1\%, as indicated above, 
implies an internal background mismatch between the open
and blocked data sets of  not more than 0.4\%.  The observed diffuse flux above
80 keV gives a count rate of about 6.7 x 10$^{-3}$ s$^{-1}$, about 1\% 
of the average background level.  The observed agreement of this
diffuse flux with that from the Medium Energy Detectors at 80--100 keV 
(Kinzer et al. 1997), where the latter detectors have a signal about 
equal to the internal background, and which is therefore very reliable,
shows that this limit of 0.4 percent is not too low at these higher
energies.  The flat spectrum of the internal background insures that
background estimation errors will have a completely negligible effect
at all lower energies, where the real strength of this measurement lies.

\subsection{Variation of Detector Gain}

The electron optics of the photomultiplier tube 
were insufficiently shielded from effects of the geomagnetic field, and 
therefore
changed with orientation, resulting
in gain variations as large as 20\% peak-to-peak, with an RMS
of the order of 5\%.  This gain variability was laboriously
modeled in detail (Jung 1986)  and corrected in the data 
so that the net instantaneous
gain error was between one and 2\% rms.  The propagation of this
error into the average spectrum over about 135 rotations of the
spacecraft reduced the net effect by another order of magnitude, making
it completely negligible.

\subsection{Energy Calibration}

Relative energy calibration was based on preflight measurements of the
differential channel width for detector pulses.  Absolute energy
calibration, required by a sudden change post-launch and a slow drift
thereafter, was monitored using two discrete features of the internal 
background.  The primary calibrator was a K-capture line of I$^{125}$, which 
produces a gamma ray of 35 keV followed by a prompt decay, followed by K and L
X-rays from the I$^{125}$ daughter, for a total of 67 keV.  Differential
response of the NaI scintillator produced light equivalent to a
single photon emitted at 62.7 keV, based on ground calibrations.  Measurement
of bright sources such as the Crab Nebula (Jung 1989) and Cyg X-1
(Nolan \& Matteson 1982) using this gain calibration produced smooth
spectra, but a variation of the formal gain value
by only a few percent from this produced an artifact at 40 keV in
each of these sources.  Our secondary calibration line, a blend of
Iodine and Tellurium K X-rays, was useful only as confirmation of
the prime calibration, because of the unknown and possibly variable 
mix of the two species.  
The background spectrum in Figure 1 shows the features used to determine the 
energy calibration.  The effective energy resolution in orbit was about 15 keV 
FWHM at 60 keV.

\begin{figure}[tb]
\centerline{\epsfig{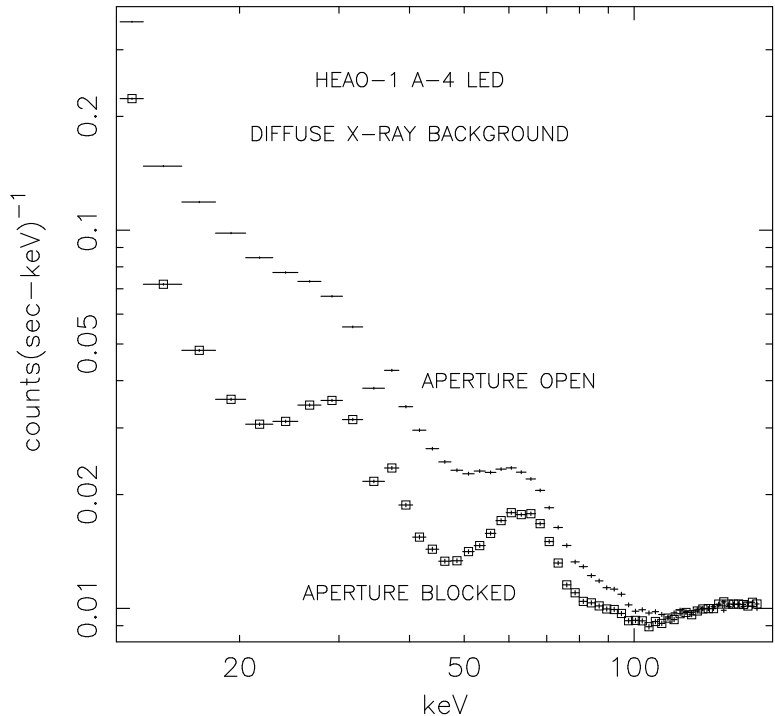}}
\caption{\label{total_net} The differential counting rates obtained by the 
Low Energy Detectors on the UCSD/MIT Hard X-ray and Gamma-Ray Instrument on 
the High Energy Astronomical Observatory--1.  The difference between the
rates with the detector blocked with an active shutter and unblocked when
looking at the sky well above the horizon is due to the  diffuse component at 
cosmic X-rays.  The rates are averaged over similar ranges of B,L  magnetic 
coordinates.  The artifact at about 32 keV is due to an energy-loss anomaly in 
NaI near the K-edge.  The diffuse flux is well above the detector background 
to at least 100 keV.}
\end{figure}

\subsection{Emission from Blocking Crystal}

The open minus blocked difference spectrum initially showed a strong
deficit near 30 keV, the effective energy of the K X-ray blend from
excited Iodine and Tellurium isotopes in the detector material.
This deficit, and its identification as K X-rays, was traced to the
blocking crystal, whose material also undergoes spallation by 
cosmic rays, followed in some cases by K-capture decay of the daughter,
with a high-energy gamma that escapes the blocking crystal, and K
radiation that produces a count in the detector.  While this process
is difficult to calculate, it was easier and more reliable to measure
the effect in earth-looking data.  We make the reasonable assumption
that the earth's secondary X-ray spectrum is featureless near 30 keV.

\section{Results}

The average counting rates of the selected Low Energy Detector (LED),
after correction for gain variations, are shown in Figure 1 for
both the blocked and unblocked data.  The sky data taken when the
beam was above the horizon, and the blocked data is averaged over all zenith 
angles.  These rates correspond
to an average L of 1.17 and an average B of 0.30.
As indicated previously the diffuse sky component is dominant at the
lower energies, and is a small fraction of the average background at above
100 keV.  Except for small corrections, the difference of these two curves 
is the rate due to diffuse hard X-rays.  

The resultant sky flux in units proportional to  $\nu F_{\nu}$ is shown
in Figure 2.  The data here are corrected for the geometry factor, 3.0
cm$^2$-sr, and for the energy response matrix.  The latter has been determined 
from a combination of direct pre-launch 
measurements, and Monte-Carlo calculations.
At these low energies, photoelectron absorption in the thin
NaI detector is the primary interaction, so a simple efficiency correction 
applies over most of the energy range.  The sky flux is shown averaged over 
many PHA channels widths, comparable to the measured energy resolution
of 15 keV at 59 keV.  Selecting energy widths of approximately 
constant ratio helps to keep the statistical significance of the plotted
channels comparable on a log-log plot.

\begin{figure}[tb]
\centerline{\epsfig{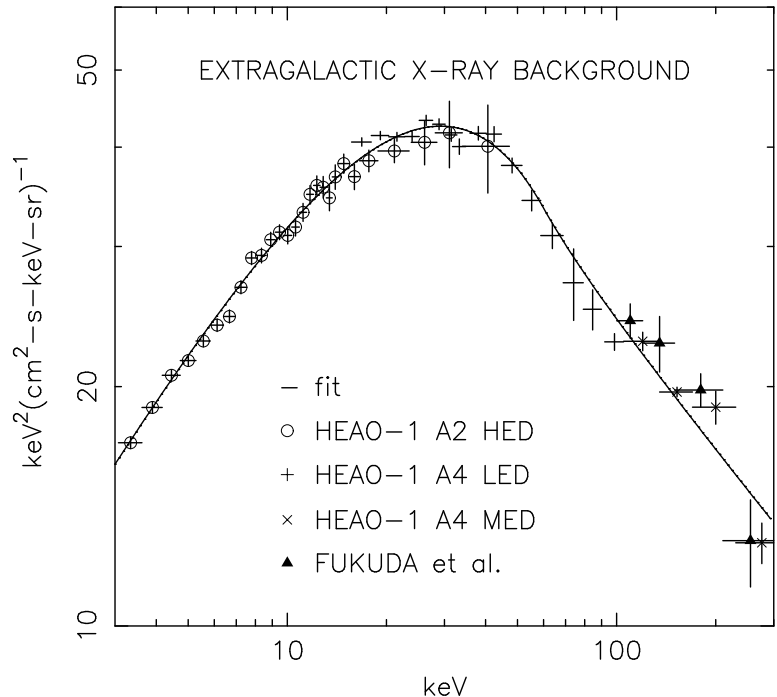}}
\caption{\label{hump}The corrected photon spectrum of the diffuse 
component measured 
with the HEAO--1 by several different detectors, compared with other data in 
this energy range, expressed as spectral intensity per logarithmic energy unit
dI/d(log E).  The HEAO A4 Low Energy Detectors join smoothly to 
other data at both higher and lower energies, and are in agreement with 
balloon data, for which that of Fukuda et al. (1975) has been chosen as 
representative.  The fit is shown as a curved line.  The data from HEAO-1
A2 were taken from High-Energy Detector no. 1 (E. Boldt, private 
communication).  These A2 points may reflect minor artifacts of the spectral 
inversion.} 
\end{figure}

Also shown in Figure 2 are the results obtained by a number of other
experiments. The HEAO--A2 instrument (Marshall et al. 1980), which produced
the most significant result on diffuse fluxes in 
the 3--45 keV range, overlaps and joins smoothly 
with the LED results.  
The LED data also join  smoothly at the higher
end to the data obtained from Medium Energy Detectors (MEDs) 
(Kinzer et al. 1997).  Data obtained by a number of 
balloon and space
experiments (Kinzer, Johnson \& Kurfess 1978; Fukuda et al.
1975)  are also shown for comparison.  As discussed in Kinzer et 
al. (1997) it is significant that data in this range obtained with a number of 
different experimental techniques and in various radiation environments are in 
agreement within statistical and systematic uncertainties.  We conclude the 
diffuse hard X-ray background is well determined in the range 
3~$<$~E~ $<$~500 keV.

\section{Total Diffuse Spectrum}


With these and other recent results, it is now possible to define the 
spectrum over the entire observed energy range above 3 keV, and to generate 
an empirical analytic fit to this spectrum.  Figure 3 shows selected  data 
presented on an intensity scale, which is more useful 
for theoretical comparisons than the photon scale.

\begin{figure}[tb]
\centerline{\epsfig{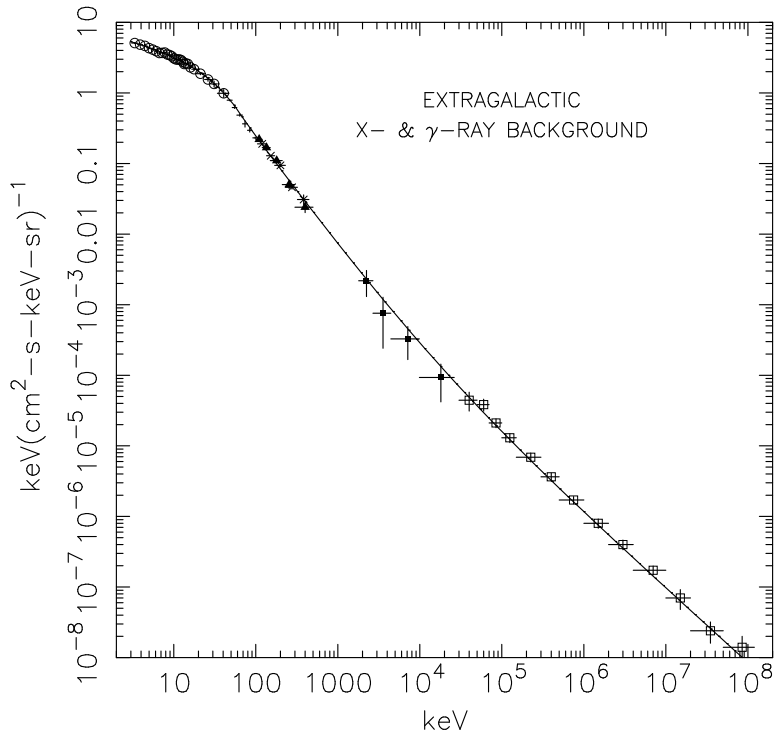}}
\caption{\label{full_range} Selected recent results on the intensity spectrum 
of the diffuse cosmic component over the 3 keV to 100 GeV range.  The results 
are fitted to simple empirical exponential and power-law functions.  The 
reduced $\chi^2$ of the fit is about 1.3, over almost eight decades of photon 
energy.  Various source classes and physical processes are postulated to 
dominate in different spectral ranges.  Data below 500 keV were chosen and 
marked as in Figure 2.  Comptel  and Egret data are marked with filled and 
open squares, respectively.}
\end{figure}

The lower energy data ($<$ 500 keV) is shown again, converted to intensities.
The COMPTEL results in the 0.8~$<$~E~$<$~30 MeV range, which fail 
to establish the ``MeV bump'' (Kappadath et al. 1996; Kinzer et al. 1997) are 
also  shown.  We do not plot the Apollo 15/16 results 
which overlap the HEAO and the COMPTEL work since uncertainty in the correction
for induced background (Trombka et al. 1977) in detectors operating over
the 0.5~$<$~E~$<$~10~MeV range has almost certainly caused the artifact 
resulting in the ``MeV bump''.  
We have also discarded other reported results from scintillators above 500 keV
as unreliable.
At higher energies, the earlier results obtained
on SAS--2 are in substantial agreement with the more definitive results
obtained over an extended energy range by EGRET on CGRO (Kniffen et al. 1996).

Since the various results in Figure 3 obtained with many instruments
and techniques over eight decades in energy appear to join smoothly, it is 
possible to empirically fit an analytic function to the data from the
full energy range.
Such a function had been developed previously by Gruber (1962), based on the 
data as it was available at the time.

The present data and selected earlier data, all shown in Figure 3, have been 
empirically fit to a combination of exponential and power law functions, 
operating over different energy ranges.  Criteria are that the functions 
join smoothly in first and second order at the break point, and that 
$\chi$$^2$ be minimized.  Such a function is:

\begin{center}
\begin{tabular}{lll}\\
3--60 keV: & 7.877 E$^{-0.29}$ e$^{-E/41.13}$ & keV/keV-cm$^2$-sec-sr\\
$>$ 60 keV: & 0.0259 (E/60)$^{-5.5}$ & \\
& + 0.504 (E/60)$^{-1.58}$ & keV/keV-cm$^2$-sec-sr \\
& + 0.0288 (E/60)$^{-1.05}$ 
\end{tabular}
\end{center}

Overall, the reduced $\chi$$^2$ is about 1.3, which may be regarded
as an excellent fit, considering the data used for the fitting was 
obtained from five  different instruments.  The function is shown as a
solid line in Figures 2 and 3.

The function below 60 keV is that introduced by Boldt (1987, 1988, 1989, 1992)
as an excellent fit to the HEAO-1 A2 data,
but has slightly different values for the normalization and e-folding energy, 
reflecting, of course, the fit to a different and larger data set consisting
of the A4 LED data from HEAO-1 and HEAO-1 A2 data from the High-Energy
Detector no. 1 (E. Boldt, private communication), which was independent
of the set analyzed by Marshall et al. (1980).  This 
lower-energy fit with the present best-fit values was first reported by Gruber 
(1992).  Boldt (1988) and Holt (1992) have both emphasized that the two 
spectral parameters, index and e-fold energy, of this function are particularly
revealing for characterizing the residual CXB spectra obtained when
subtracting various foreground components.

Above 60 keV selected data sets included the HEAO A4 (LED and MED), balloon, 
COMPTEL and EGRET data.  The fit required the sum of three power laws, 
the flattest of which largely characterizes the EGRET observations (it
ignores a likely ``ripple'' at 70 MeV), and the next steeper, with index 1.58,
may be said to represent the spectrum between 70 keV and an MeV.  
The steepest component, with index 5.5, is almost certainly only a numerical
necessity for matching to the lower-energy spectrum and its derivative, and 
represents nothing physical.

The three main functional components may possibly be identified with
separate physical components.
If the flat EGRET component continues unbroken to much lower energies, 
and the rollover at tens of keV is an actual cutoff for the lower
energy component, then the index 1.58 power law characterizes a separate
component dominant at hundreds of keV.

Given the lower-energy spectral form, the maximum in $\nu F_{\nu}$ (see Figure
2)  of 42.6 kev(sec\,cm$^2$\,sr)$^{-1}$ occurs at 29.3 keV, very close to the 
values of 41.3 kev(sec\,cm$^2$\,sr)$^{-1}$ and 28.4 keV, respectively, for A2 
data alone (E. Boldt, private communication), indicating that the results
from the HEAO-1 experiments are robust, both with respect to normalization and
spectral shape.

\section{Discussion}

The final analysis of the HEAO A4 Low Energy Detectors presented here, and that 
of the Medium Energy Detectors presently earlier (Kinzer et al. 1997) have 
provided a completely consistent set of measurements of the diffuse component 
of cosmic X-ray over the range 13 $<$ E $<$ 400 keV.  These data join
smoothly to other recent data obtained at higher energies by the COMPTEL
(Kappadath et al. 1996) and EGRET (Sreekumar et al. 1998) instruments on the 
Compton Gamma Ray Observatory.  ASCA and ROSAT (Chen, Fabian \& Gendreau 1997) 
have presented new results in the 0.1--7 keV band.  These data, and those of 
the HEAO A4, agree well with those previously presented in the 3--45 keV range
from the HEAO A2 instrument (Marshall et al. 1980).  Only in the range 
$\sim$300 $<$ E $<$ $\sim$1 MeV is a set of new or confirming data missing; the 
earlier Apollo 15/16 data in this range now being suspect.  To 
obtain an accurate, 
definitive spectrum in this range will require a new instrumental concept, 
since  instruments designed for this range, such as OSSE on CGRO (Kurfess 1996) 
and the spectrometer SPI to be launched on INTEGRAL (Mandrou et al. 1997) have 
relatively narrow apertures and high background, more optimized for discrete 
source studies.

The excellent (reduced  $\chi^2$ = 1.3) of a simple exponential at lower 
energies and three summed power law functions above 60 keV to our selected 
set of data over the entire 3 keV to 100 GeV range 
can only be described as remarkable.  This is particularly so, 
considering that different discrete source classes producing X- and gamma-rays 
by different mechanisms are certainly operating in the different energy 
ranges.  It seems at present that a truly cosmological origin for the 
``diffuse'' cosmic component in unlikely in any energy range, and that
the integrated effects of various evolving classes of discrete sources are 
sufficient to explain the phenomenon.  Small discontinuities and inflections 
expected in the combined spectra due to various physical processes predicted 
by The et al. (1993) have not been observed with high 
resolution instruments (Barthelmy et al. 1996).  The data presented here 
is of low 
resolution ($\Delta$E/E $\geq$ 0.1), or is averaged over wide bands
($\Delta$E/E $\geq$ 0.2), precluding  searches for narrow band
discontinuities or inflection phenomena.

Even so, it requires a rather unique combination of power law quasar X-ray 
spectra and absorbed Seyfert II's to produce the very smooth exponential in the 
3--60 keV range.  Such a class has been postulated by Madau et al. (1994), 
Comastri et al. (1995) and Zdziarski (1996).  Studying a large red-shifted
class of absorbed Seyferts II's to determine the distributions of low-and
high-energy cutoffs is crucial to resolving this problem.  A similar problem 
exists for the integrated effect of Supernova Ia's to cosmological distances to 
explain the diffuse spectrum in the MeV range (The et al. 1993).  Here effects 
due to line emission are expected to produce discontinuities; as indicated
above such effects have been searched for and not found.
However, the recent discovery of a class of ``MeV blazars'', with emission 
concentrated near 1 MeV (Collmar, private communication 1998) may provide an 
alternative to the supernova component.

To make further progress on understanding the diffuse component of cosmic X- 
and gamma-rays therefore requires advances on two observational fronts.  First,
high sensitivity, high resolution class studies of postulated source components 
are needed to determine the luminosity function of the various spectral types.
Second, high resolution, low background instruments specifically designed to 
measure the diffuse cosmic flux are required for precise determination of 
spectral features, particularly in the range about 10 keV to 1 MeV, where the 
various postulated components join, and where phenomena due to discrete
lines may be operative.

\section{Acknowledgements}

We acknowledge the contribution of many students, colleagues and technical 
support personnel to the HEAO program. We have received many useful
comments from Elihu Boldt, R. E. Lingenfelter and R. E. Rothschild.
G.V. Jung was a Ph.D. student at UCSD during the course of this work.
This work 
was supported by NASA under contract NAS8--27974 and grants NAGW--449.

\newpage

\begin{center}
{\bf REFERENCES}
\end{center}

\begin{description}
\item[] Barber, C. R. \& Warwick, R. S. 1994, MNRAS, 267, 270
\item[] Barcons, X., Fabian, A. C., \& Rees, M. J. 1991, Nature, 350, 685
\item[] Barthelmy, S. D., Naya, J. E. Gehrels, N., Parsons, A., Teegarden, B., 
Tueller, J., Bartlett, L. M., \& Leventhal, M. 1996, {\em BAAS}, {\bf    },
\item[] Boldt, E. 1987, Phys. Reports, 146, 215.
\item[] Boldt, E. 1988, in Physics of Neutron Stars and Black Holes, ed. Y.
Tanaka (Tokyo: Universal Academy press), 342.
\item[] Boldt, E. 1989, in X-Ray Astronomy, 2. ESA SP-296 (Noordwijk:
ESTEC), 797.
\item[] Boldt, E. 1992, in The X-Ray Background, ed. X. Barcons \& A. C.
Fabian (Cambridge: Cambridge Univ. Press), 116.
\item[] Briggs, M. S. 1992, Ph.D. thesis, Univ. California, San Diego
\item[] Chen, L.W., Fabian, A. C., \& Gendreau, K. C. 1997, MNRAS, 285, Issue 
3, 449, astro-ph/9711083
\item[] Comastri, A., Setti, G., Zamorani, G. \& Hasinger, G. 1995, A\&A, 296, 
1.
\item[] Fabian, A. C. \& Barcons, X. 1992, Ann. Rev. Astron. \& Astrophys.,
30, 429
\item[] Fichtel, C. A., Simpson, G. A., \& Thompson, D. J. 1978,
ApJ, 222, 833
\item[] Fukada, Y., Hayakawa, S., Kasahara, I., Makino, F., \& Tanaka, Y. 1975, 
Nature, 254, 398
\item[] Gendreau, K. C. et al. 1995, PASJ, 47, L5
\item[] Giacconi, R., Gursky, H., Paolini, R., \& Rossi, B. 1962, Phys. Rev. 
Lett., 9, 439
\item[] Gruber, D. E. 1974, Ph.D. thesis, Univ. of CA at San Diego
\item[] Gruber, D. E., Jung, G. V., \& Matteson, J. L. 1989, in High Energy 
Radiations in Space, ed. A. C. Restor, Jr. \& J. I. Trombka (AIP: New York), 
232
\item[] Gruber, D. E. 1992, in The X-Ray Background, ed. X. Barcons \& A. C. 
Fabian (Cambridge: Cambridge Univ. Press), 46
\item[] Hasinger, G. 1996, A\&A Supp. Series, 120, 607
\item[] Hasinger, G. et al. 1998, A\&A, 329, 482
\item[] Holt, S. 1992, in The X-Ray Background, ed. X. Barcons \& A. C.
Fabian (Cambridge: Cambridge Univ. Press), 33
\item[] Horstman, H. M., Cavallo, G., \& Moretti-Horstman, E. 1975, Nuovo 
Cimento, 5, 255
\item[] Jung, G. V. 1986, Ph.D. thesis, Univ. of CA at San Diego
\item[] Jung, G. V. 1989, ApJ, 338, 972
\item[] Kappadath, S. C., et al. 1995, in Proc. 24th Intl. Cosmic-Ray Conf. 
(Rome), Vol. 2, 230
\item[] Kappadath, S.C., et al. 1996, A\&A Supp. Series, 120, 619
\item[] Kinzer, R. L., Johnson, W. N., \& Kurfess, J. D. 1978, ApJ, 222, 370
\item[] Kinzer, R. L., Jung, G. V., Gruber, D. E.,
Matteson, J. L. \& Peterson, L. E. 1997, ApJ, 475, 361
\item[] Kinzer, R. L., Johnson, W. N., \& Kurfess, J. D. 1978, ApJ, 222, 370
\item[] Kraushaar, W. L. et al. 1972, ApJ, 177, 341
\item[] Kurfess, J. D. 1996, A\&A Supp., 120, 5
\item[] Kniffen, D. A., et al. 1996, A\&A Supp. Series, 120, 615.
\item[] Levine, A. M. et al. 1984, ApJ Supp., 54, 581
\item[] Madau, P., Ghisellini, G., \& Fabian, A. 1993,
ApJ, 410, L7
\item[] Madau, P., Ghisellini, G., \& Fabian, A. 1994, MNRAS, 270, L17
\item[] Mandrou, P. et al. 1997, Proc. 2nd INTEGRAL Workshop (ESA Pub.), 591
\item[] Marshall, F. E., Boldt, E. A., Holt, S. S., Miller, R. B.,
Mushotzky, R. F., Rose, L. A., Rothschild, R. E. \& Serlemitsos, P. J. 1980,
ApJ, 235, 4
\item[] Matteson, J. L. 1978, in Proc. AIAA 16th Aerospace Science Meeting, 
78--35, 1
\item[] McIlwain, C E. 1961, J. Geophys. Res., 66, 3681
\item[] Metzger, A. E., Anderson, E. C., Van Dilla, M. A., \& Arnold, J. R. 
1964, Nature, 204, 766
\item[] Naya, J. E., Barthelmy, S. D., Bartlett, L. M., Gehrels, N., Parsons, 
A. et al. 1998, ApJ, 499, L169, astro-ph/9804074 
\item[] Nolan, P. L. \& Matteson, J. L. 1983, ApJ, 265, 389
\item[] Rothschild, R. E., Mushotzky, R. F., Baity, W. A., Gruber, D. E.,
Matteson, J. L., \& Peterson, L. E. 1983, ApJ, 269, 423
\item[] Sreekumar, P. et al. 1998, ApJ, 494, 523, astro-ph/970925
\item[] Stecker, F. W. \& Salamon, M. H. 1996, ApJ, 464, 600, astro-ph/9609102
\item[] The, L.-S., Leising, M. D., \& Clayton, D. D. 1993, ApJ, 403, 32
\item[] Trombka, J. I. et al. 1977, ApJ, 212, 925
\item[] Zdziarski, A. A. 1996, MNRAS, 281, L9
\end{description}
\newpage

\begin{center}
TABLE 1\\
Detector Properties
\end{center}

\begin{tabular}{cccccc} \\ \hline \hline
Detector & Number & Energy & Area & FOV & Geometry \\
& & keV & cm$^2$ & degrees & cm$^2$-ster \\ 
& & (nominal) & & (FWHM) & \\ \hline \hline 
\\
Low Energy (LED) & 2 & 13--180 & 103 ea & 1.7$\times$20 & 3.0\\
Medium Energy (MED) & 4 & 80--2100 & 42 ea & 17 & 3.97 \\
High Energy (HED) & 1 & 150--10000 & 120 & 30 & 100 \\ \hline
\end{tabular}

\end{document}